\documentclass[12pt,preprint]{aastex}

\usepackage{epsf}
\usepackage{graphics}
\input{psfig}

\newcommand{\msun}{$M_{\odot}$}
\newcommand{\dM}{$\dot M$}
\newcommand{\dm}{$\dot m$}

\newcommand{\dMedd}{$\dot M_{\rm Edd}$}

\newcommand{\sigo}{$\sigma_{\rm [OIII]}$}
\newcommand{\sigs}{$\sigma_{*}$}
\newcommand{\sig}{$\sigma$}
\newcommand{\mbh}{$M_{\rm BH}$}
\newcommand{\chandra}{{\it Chandra}~}
\newcommand{\etal}{{\it et al.}~}
\shorttitle{NLS1s \& the \mbh--\sig\ relation}
\shortauthors{Grupe}

\begin{document}


\def \charthoffset {\hspace{0.2cm}} \def \charthsep {\hspace{0.3cm}}
\def \chartvsepcap {\vspace{0.3cm}}
\def \chartvsep {\vspace{0.1cm}}
\newcommand{\putchartb}[1]{\clipfig{/home/halley/dgrupe/ps/#1}{85}{20}{0}{275}{192}}
\newcommand{\putchartc}[1]{\clipfig{/home/halley/dgrupe/ps/#1}{55}{33}{19}{275}{195}}
\newcommand{\chartlineb}[2]{\parbox[t]{18cm}{\noindent\charthoffset\putchartb{#1}\charthsep\putchartb{#2}\chartvsep}}

\newcommand{\chartlinec}[2]{\parbox[t]{18cm}{\noindent\charthoffset\putchartc{#1}\charthsep\putchartc{#2}\chartvsep}}


\title{The Locus of Highly Accreting AGNs on the \mbh--\sig\ Plane:
 Selections, Limitations, and  Implications}


\author{Smita Mathur \& Dirk Grupe\altaffilmark{1}}
\affil{Astronomy Department, Ohio State University,
    140 W. 18th Ave., Columbus, OH-43210, U.S.A.}
\email{smita@astronomy.ohio-state.edu}


\altaffiltext{1}{Department of Astronomy and Astrophysics, Pennsylvania
 State University, University park, PA 16802}


\begin{abstract}
We re-examine the locus of narrow line Seyfert 1 galaxies on the
\mbh--\sig\ (black hole mass--bulge velocity dispersion) plane in the
light of the results from large new optically selected samples. We find
that (1) soft X-ray selected NLS1s have a lower ratio of BH mass to
$\sigma^{4}_{[OIII]}$ than broad line Seyfert 1 galaxies; this remains a
robust statistical result contrary to recent claims otherwise; (2)
optically selected NLS1s have systematically lower Eddington luminosity
ratio compared to X-ray selected NLS1s; and (3) as a result, the locus
of NLS1s on the \mbh--\sig\ plane is affected by selection effects. We
argue that there is no single explanation for the origin of the
\mbh--\sig\ relation; instead tracks of galaxies on the \mbh--\sig\
plane differ with redshift, consistent with the downsizing of AGN
activity. If these results at face value are incorrect, then the data
imply that AGNs with high Eddington accretion reside preferentially in
relatively late type galaxies at the present epoch, perhaps a more
interesting result and a challenge to theoretical models.
\end{abstract}


\section{Introduction}

How do black holes (BHs) form, how do they grow, when do they become
``active'' as in quasars and low redshift active galactic nuclei (AGNs),
what is the accretion history of BHs, how does it relate to their active
phase and how does all this relate to the formation and evolution of
galaxies? These questions have received considerable attention in the
literature in the past five years or so, at least in part due to the
discovery of the \mbh--\sig\ relation \cite{geb00a, ferr00, merr01}. The
tight correlation between BH mass and the bulge velocity dispersion of
its host galaxy implies that the ultimate fate of the two is linked. It
is not obvious, however, how galaxies and their BHs find their way onto
the \mbh--\sig\ relation. A large number of theoretical models attempt
to explain the observed correlation with a variety of physical
processes; regulation of the bulge growth by the feedback from the
active BH appears to be a popular one (e.g. King \& Pounds 2003, Hopkins
et al. 2005; see Mathur \& Grupe 2004 for a more extensive list of
references).



Do all BHs follow the same track on the \mbh--\sig\ plane or does it
depend on BH mass, redshift, galaxy properties or any other parameter? 
It is of interest, therefore, to find loci of high redshift AGNs on the
\mbh-- \sig\ plane. Rix \etal\ (1999) find that z$\sim 2$ quasars have
higher BH mass to host galaxy stellar mass ratio compared to that at
z=0; as such they would lie above the Tremaine \etal fit \cite{tre02} to
the \mbh--\sig\ relation. This implies that high redshift luminous BHs
grew fast while their host galaxies were still in the process of
assembling. At low redshift, on the other hand, galaxies presumably have
finished growing while black holes accreting at the Eddington rate will
e-fold their mass in a Salpeter time. Finding the locus of highly
accreting BHs on the \mbh--\sig\ plane is also of interest, therefore,
to understand the origin of the \mbh\--\sig\ relation.  In the local
universe, a class of Seyfert galaxies called the narrow line Seyfert 1
galaxies (NLS1s) are known to be highly accreting AGNs (Pounds et
al. 1995). The locus of NLS1s on the \mbh--\sig\ plane is therefore of
considerable interest and is the subject of this paper. While there may
not be a single answer to the origin of the \mbh--\sig\ relation,
e.g. it may differ with redshift, NLS1s promise to provide at least a
piece of the puzzle.

\section{NLS1s and the \mbh--\sig\ relation}

Methods that work well for measuring BH masses in normal galaxies, such
as gas dynamics and stellar dynamics, do not work well for active galaxies;
the glare of the active nucleus makes it difficult to use these
techniques.  For AGNs, reverberation mapping provides a powerful
technique to measure BH masses and has been employed successfully on
nearby Seyfert galaxies (Peterson 1993). Reverberation mapping, however, is
time intensive, so reliable BH mass measurements have been made of only 
a small number of AGNs. Based on the reverberation mapped AGNs, Kaspi \etal
determined an empirical relation between \mbh\ and the width of the
H$\beta$ emission line \& optical continuum luminosity
\cite{kas00}. This simple and well calibrated relation can be easily
used to estimate BH masses in a large number of AGNs and has been used
by a number of authors \cite{mclure02,shi03}. Measuring
\sig\ poses a bigger problem, because the strong AGN light washes out
underlying stellar absorption lines. The width of the narrow [OIII]
emission line is often used as a surrogate for the bulge \sig. It was
noted by Whittle (1992) that the kinematics of the narrow line region of
AGNs is governed by the gravitational potential of the host galaxy
bulge, and not by the nuclear BH. Nelson \& Whittle (1995a,b) found
that FWHM([OIII]) is correlated with bulge \sig. The outliers in the
correlation were sources with strong radio jets; since most AGNs are
radio-quiet, without strong jets, the use of FWHM([OIII]) as a surrogate for
bulge \sig\ appeared to be reasonable.

Mathur \etal (2001) were the first to place NLS1s on the \mbh--\sig\
plane and found that they do not follow the same relation as broad line
Seyfert 1s (BLS1s) and normal galaxies. They used X-ray spectral energy
distribution to estimate the BH masses and [OIII] widths to estimate
\sig. Using a complete sample of soft X-ray selected AGNs, Grupe \&
Mathur (2004, Paper I hereafter) confirmed the above
result. Specifically, they found that for a given \sig, NLS1s have
smaller BH masses compared to BLS1s. Their BLS1 and NLS1 samples spanned
the same range in luminosity, so were well matched. In paper I, the
H$\beta$ width was used to measure BH mass and again, [OIII] widths were
used to estimate \sig. Below we consider limitations of, controversies
about, and implications of the above result.

\subsection{Limitations}

Paper I discusses the limitations of methods to estimate \mbh\ and
\sig\ in detail. Here we reiterate a few for the sake of completeness
and emphasize a few more. There are many sources of error in using the
Kaspi \etal (2000) relation to estimate BH masses: (1) this relation is
calibrated on reverberation mapping measurements of \mbh, which itself
is uncertain by a factor of few because of the unknown geometry of the
broad emission line region. (2) While there are a few NLS1s in the broad
line region radius -- luminosity correlation in Peterson \etal (2000),
their sample is not large enough to cover the observed range of \mbh\ in
our samples. (3) Extrapolation of the Kaspi et al. relation to higher or
lower masses would introduce an additional source of error in BH mass
estimates.  As a result, errors on individual \mbh\ values are large.

The errors on individual \sig\ values are also large for various
reasons. The correlation between FWHM([OIII]) and \sig\ \cite{nel00} has
a large scatter which produces one source of error. Secondly, [OIII]
lines often show some blue asymmetry, which would overestimate
FWHM([OIII]) of the core component. We corrected for the asymmetry in
the [OIII] line profile in Paper I; even so, the errors on
\sig\ remain large.

Because of large errors on individual \mbh\ and \sig\ measurements, we
have emphasized in Paper I that the results  are statistical
in nature and are robust for determining the aggregate properties of the
samples. These, however, are the results at face-value. Confirmation of
these results require more accurate measurements of \mbh\ and \sig.

\subsection{Controversies}

The aforementioned results, however, are controversial. While the use of
FWHM(H$\beta$) as a surrogate for \mbh\ is well accepted, the same
cannot be said about FWHM([OIII]) as a surrogate for \sig. Perhaps the
most important issue regarding the use of \sigo\ as a surrogate for
\sig\ was highlighted recently by Greene \& Ho (2005, GH05
hereafter), which we discuss below.

Using a large sample of narrow line AGNs\footnote{Please note that these
narrow line AGNs are {\it not} NLS1s. These are AGNs in which only
narrow lines from the narrow lines region of AGNs are visible while the
strong nuclear continuum and the broad line region are hidden. These are
traditionally referred to as type 2 AGNs.} selected from the Sloan
Digital Sky Survey (SDSS) GH05 compared the bulge velocity dispersion
\sigs\ measured using the underlying host galaxy spectrum and \sigo\
using the narrow [OIII] emission line. This was an important study as it
contained a very large sample, compared to the original work of Nelson
\& Whittle (1995, 1996). Moreover, the sample selection, data reduction and
analysis were performed in a homogeneous way. Based on this comparison,
GH05 concluded that indeed, the kinematics of the NLR gas is dominated
by the bulge gravity; the widths of the low ionization NLR emission
lines such [SII] and [OII] track the stellar velocity dispersion in the
mean, albeit with substantial scatter. As such, they may be used as a
proxy for \sigs. On the other hand, the width of the [OIII] line is
significantly broader than \sigs, so \sigo\ cannot be used as a proxy
for \sigs. However, when the blue asymmetric wing of [OIII] is removed,
the width of the core component does track \sigs. This is an important
conclusion because it validates the use of
\sigo\ (after removing the blue wing) as a surrogate for \sigs\ in
previous studies and will help many future studies. Moreover, as
mentioned above, the main result of Paper I (that NLS1s have lower mass
BHs than BLS1s for a given \sig) was derived after removing the blue
asymmetry of [OIII] lines, and is thus a statistically robust result
(see below). Nonetheless, GH05 point out that the scatter around the
\sigo--\sigs\ relation is large, even after removing the blue wing, so
\sigo\ of the core component should be used as a proxy for
\sigs\ only in a statistical sense, as done and emphasized in Paper
I. Boroson (2003) also pointed out the same using SDSS early release
data, viz. the scatter in \sigo\ around the \mbh--\sig\ relation is
large and therefore \sigo\ should be used as a surrogate for \sigs\ only
in a statistical sense. The Nelson \& Whittle studies (1995, 1996) had
one advantage over GH05; they could effectively identify the outliers in
the \sigo--\sigs\ correlation with disturbed galaxies and/or powerful
linear radio sources. Given the SDSS data quality and the sensitivity of
the FIRST survey used by GH05, such identifications could not be
made. One should also keep in mind that the GH05 study is based on type
2 AGNs; while it is reasonable to extend it to type 1 AGNs if
orientation is the only difference between the two types, it may not be
so.

Greene \& Ho (2005) then go on to find secondary drivers of the
deviations of \sigo\ from \sigs, parameterized with
$\Delta\sigma\equiv\log$\sigo$-\log$\sigs. They consider host galaxy
morphology, local environment, star formation rate, bulge velocity
dispersion, radio power, AGN luminosity, and the ratio of bolometric to
Eddington luminosity (L$_{\rm bol}/$L$_{\rm Edd}$) as possible secondary
drivers. They do not find any strong correlation between $\Delta\sigma$
and any of these parameters (though note the caveat above) except
L$_{\rm bol}/$L$_{\rm Edd}$. There appears to be a mild but systematic
trend of higher $\Delta\sigma$ in objects with higher L$_{\rm
bol}/$L$_{\rm Edd}$ (formal Spearman rank correlation coefficient of
0.46, with a probability of chance correlation P$<0.0001$) as shown in
equation 3 of GH05, which is

\begin{equation}
\Delta\sigma = (0.072\pm0.005)\log L_{bol}/L_{Edd} + (0.080\pm0.005)
\end{equation}

for \sigo $\equiv$FWHM([OIII])$/2.35$. This led GH05 to conclude that
\sigs\ is overestimated in objects with high L$_{\rm bol}/$L$_{\rm
Edd}$, such as NLS1s and high redshift quasars (Paper I, Shields et
al. 2003).

Given the implications of these results (\S 3), it is important to
establish whether $\Delta\sigma$ is truly a function of L$_{\rm
bol}/$L$_{\rm Edd}$. GH05 have calculated L$_{\rm bol}$ using observed
L$_{\rm [OIII]}$ and a bolometric correction factor. Since their SDSS
sample consists of narrow line AGNs, broad H$\beta$ lines are not
observed, so there is no direct handle on BH mass, and so on L$_{\rm
Edd}$. GH05 use the  \mbh--\sigs\ relation of Tremaine et al
(2003) to derive \mbh\ and so L$_{\rm Edd}$ from the observed values of
\sigs. Thus a function of \sigs\ is compared to $\Delta\sigma$, which in
itself is a function of \sigs\ suggesting that the correlation between
$\Delta\sigma$ and L$_{\rm bol}/$L$_{\rm Edd}$ claimed by GH05 may 
 be a result of a circular argument. To investigate further whether
this is indeed the case, we re-wrote the above correlation equation in
terms of the actual observed quantities. We find that it translates to:

\begin{equation}
\log \sigma_{[OIII]} = A \log L_{[OIII]} + B \log \sigma_* + C
\end{equation}

where numerical values of constants A, B and C are
a result of correlations between L$_{\rm bol}$ and L$_{\rm [OIII]}$,
\mbh\ and \sigs, and equation 1. Thus we see from equation 2 that the 
actual relations underlying equation 1 are a mild correlation between
\sigo\ and L$_{\rm [OIII]}$ (with a slope A=$0.072$) and a strong correlation
between \sigo\ and \sigs (with a slope B=$0.71$), {rather than the
correlation between $\Delta\sigma$ and L$_{\rm bol}/$L$_{\rm Edd}$.}

GH05 have also noted that $\Delta\sigma$ correlates strongly with \sigs
(as can be seen from equation 2), or with \mbh, assuming \mbh--\sig\
relation. Since L$_{\rm bol}/$L$_{\rm Edd}$ depends upon BH mass, they
consider whether \mbh\ or L$_{\rm bol}/$L$_{\rm Edd}$ is the primary
driver of $\Delta\sigma$. They conclude that L$_{\rm bol}/$L$_{\rm Edd}$
is the primary physical parameter because the correlation of
$\Delta\sigma$ with this parameter is stronger (Spearman rank
correlation coefficient $= 0.46$) than that with \mbh\ (Spearman rank
correlation coefficient $= -0.32$). As shown above with equation 2, part
of the correlation with L$_{\rm bol}/$L$_{\rm Edd}$ has come about
because of the correlation of $\sigma_{[OIII]}$ with $L_{[OIII]}$ and
the rest because of the correlation of $\sigma_{[OIII]}$ with
\sigs\ (Nelson \& Whittle 1996). 

Is \sigo\ really correlated with L$_{\rm [OIII]}$? A literature search
showed that such a correlation indeed exists in the 2dF quasar sample
(Corbett et al. 2003) with the probability of a chance correlation
$P=0.005$ (see also Whittle 1985). In fact, [OIII]$\lambda 5007$ is the
only narrow line showing a correlation with luminosity in their
sample. The slope and normalization of the correlation are not given in
Corbett et al. (2003); however, they give the slope and normalization of
the correlation between [OIII] width and L$_{\rm b}$, the luminosity
estimated from the absolute photographic b$_{\rm J}$ magnitude
contributing to the emission line. It is interesting to note that the
slope of their FWHM([OIII])--L$_{\rm b}$ correlation is $0.12\pm0.043$,
consistent with A=$0.072$ in equation 2.  We also looked for this
correlation in the NLS1 sample in Paper I, using the data in Grupe et
al. (2004). Indeed, the data are consistent with a mild correlation
between \sigo\ and L$_{\rm [OIII]}$ with slope A=$0.072$. Thus it is
apparent that at least part of the claimed result of GH05, that
$\Delta\sigma$ correlates with L$_{\rm bol}/$L$_{\rm Edd}$, can be
explained in terms of the observed correlation between \sigo\ and
L$_{\rm [OIII]}$. The rest is due to the correlation of
$\sigma_{[OIII]}$ with \sigs\ (Nelson \& Whittle 1996). Note also that a
strong correlation between L$_{\rm [OIII]}$ and \sigs\ is reported by
Nelson \& Whittle (1996), which may arise from the correlations between
L$_{\rm [OIII]}$ and \sigo\ and between \sigo\ and \sigs.


We further investigated to what extent the results of Paper I would be
compromised, if at all, if equation 1 were in fact a true
correlation. There is other evidence in the literature suggesting that
the excess [OIII] width is correlated with L$_{\rm bol}/$L$_{\rm
Edd}$. Recently, Boroson (2005) has systematically studied the [OIII]
lines in a sample of 400 AGN spectra selected from the SDSS first data
release. Unlike the GH05 sample, the spectra of the Boroson sample
contain the broad H$\beta$ lines, giving direct estimates of BH masses
and so of L$_{\rm Edd}$. He finds that (1) objects with higher Eddington
ratio are more likely to have large [OIII] blueshifts; and (2) objects
with large [OIII] blueshifts have anomalously broad [OIII] emission
lines. However, these are not tight correlations; there are high L$_{\rm
bol}/$L$_{\rm Edd}$ objects that do not show blueshifts, and lower
L$_{\rm bol}/$L$_{\rm Edd}$ objects that do. Nonetheless, given the
Boroson (2005) results, we need to investigate the effect of \sigo--
L$_{\rm bol}/$L$_{\rm Edd}$ correlation on the result of Paper I.

As shown in Paper I, the BH masses of our two samples of BLS1s and NLS1s
are significantly different. If both samples followed the \mbh--\sig\
relation, \mbh $\propto\sigma^{4.02}_{[OIII]}$ (Tremaine et
al. 2003). We calculated the \mbh\ to $\sigma^{4.02}_{[OIII]}$ ratio for all
the objects in our sample, with new \sigo\ values calculated using
equation 1. These new values of \sigo\ are used in Figure 1, where we
plot the cumulative fraction for a K-S test of the distributions of
$\log$\mbh $-4.02\log$\sigo\ for the two populations of BLS1s and NLS1s
of Paper I. It can be clearly seen that the two populations are
significantly different, with the formal K-S test probability of being
drawn from the same population P$<0.001$. A Student's t-test gives the
probability of the two populations being similar to be P$<0.0001$. We
thus conclude that the \mbh\ to $\sigma^{4.02}_{[OIII]}$ ratio of NLS1s
is statistically smaller than that of BLS1s.

Since equation 1 is defined for $\Delta\sigma$ in which \sigo $\equiv$
FWHM([OIII])/2.35, that is what we used for \sigo\ in the K-S test shown
in figure 1. However, as discussed above, it is better to remove the
contribution from the blue wing of [OIII] before measuring its width. In
figure 2, we plot the cumulative fraction for a K-S test of $\log$\mbh
$-4.02\log$\sigo\ in which \sigo\ is measured after removing the blue
wing\footnote{This is similar to Figure 4 of Paper 1, except that the
statistic used is $\log$\mbh $-4.02\log$\sigo\ instead of $\log$\mbh
$-\log$\sigo\ }. The two populations are clearly different with the K-S
test probability of being drawn from the same population P$<0.001$
(t-test probability P$<0.0001$). GH05 do not give correlation equation
for $\Delta\sigma$-- L$_{\rm bol}/$L$_{\rm Edd}$ when \sigo\ is measured
after removing the blue wing.  Nonetheless, we once again apply the 
 ``correction'' of equation 1 to the \sigo\ values calculated after removing
the blue wing and perform the K-S test again. The two populations are
still significantly different, with the probability of being drawn from
the same population P$=0.003$ (t-test probability P=$0.0017$).

The above exercises demonstrate that the soft X-ray selected samples of
NLS1s and BLS1s of Paper I are statistically significantly different in
\mbh\ to $\sigma^{4.02}_{[OIII]}$ ratio, even when correlation of
$\Delta\sigma$ with L$_{\rm bol}/$L$_{\rm Edd}$, as given in equation 1,
is accounted for.

\subsection{Why some NLS1s lie close to the \mbh--\sig\ relation.}

As noted by many authors (Mathur et al. 2001, Ferrarese et al. 2001,
Grupe \& Mathur 2004, Mathur \& Grupe 2005, Barth et al. 2005), not all
NLS1s have relatively smaller BH mass for their bulge \sig. While some
values of \sig\ were estimated using \sigo\ as discussed above, some
actual measurements of \sig\ in NLS1 host galaxies also exist. Ferrarese
et al. (2001) measured \sig\ in NLS1 galaxy NGC~4051 for which BH mass
is measured using reverberation mapping (Peterson et al. 2000). They
found the source to lie close to the \mbh--\sig\ relation. This,
however, is no surprise because NGC~4051 lies close to the \mbh--\sig\
relation even in Mathur \etal (2001) in which \sigo\ was used as a proxy
for \sigs. More recently, Barth et al. (2005) measured \sig\ using Mg~b
and/or CaII stellar absorption lines for a sample of NLS1s selected from
SDSS (Greene \& Ho 2004). They also find that the measured \sig\ is not
significantly different from that expected from the \mbh--\sig\
relation. Clearly, these and similar such results are at odds with the
expectation that NLS1s have growing BHs (Mathur et al. 2001, Paper I)
and are thus young AGNs (Mathur 2000). One may argue that all objects
with accurate measurements of \sigs\ lie on the \mbh--\sig\ relation, so
again, [OIII] line widths must be at fault. However, the example of
NGC~4051, and the fact that \sigo\ after removing the blue wing does
track \sigs, all suggest that some other factor is likely involved in this
apparent contradiction.

One clue towards the reconciliation of these conflicting 
results comes from the work of Williams, Mathur \& Pogge (2004). Since
most large NLS1 samples were soft-X-ray selected, they were clearly
biased towards X-ray bright objects. To remedy this situation, Williams,
Pogge \& Mathur (2002) constructed a large, uniformly selected optical
sample of NLS1s from the SDSS early data release and found that only a
fraction of them were detected in the ROSAT All Sky Survey. They
performed follow up \chandra observations of ROSAT undetected sources and
found that NLS1s are a mixed bag. Not all of them are soft-X-ray bright
or have steep X-ray spectra indicative of high values of L$_{\rm
bol}/$L$_{\rm Edd}$. Moreover, they found that the soft-X-ray power-law
slope correlates with L$_{\rm X}/$L$_{\rm Edd}$, with flat spectrum
sources having lower L$_{\rm X}/$L$_{\rm Edd}$ (see also Grupe 2004; Lu
\& Yu 1999). This strongly suggests that not all NLS1s are highly
accreting sources and a large fraction of optically selected NLS1s falls
in this category. Indeed, Mathur \& Grupe (2005) have shown that NLS1s
in their sample which have higher L$_{\rm bol}/$L$_{\rm Edd}$ have
larger \sigo\ than those with similar BH masses, but lower L$_{\rm
bol}/$L$_{\rm Edd}$, and are thus likely to be growing.

We started to find the locus of NLS1s on the \mbh--\sig\ plane because
NLS1s {\it as a class} were thought to have large accretion rates
relative to Eddington (\dm $\equiv$\dM/\dMedd) leading to large L$_{\rm
bol}/$L$_{\rm Edd}$ compared to BLS1s. The above
results, however, show that not all NLS1s have large \dm\ and indicate
that  sample selection methods strongly influence the results. To
investigate this further, we compared the distributions of L$_{\rm
bol}/$L$_{\rm Edd}$ for three samples: the soft-X-ray selected sample of
NLS1s from Paper 1, the optically selected sample of NLS1s from Greene
\& Ho (2004), and the BLS1 sample from Paper I. The L$_{\rm
bol}/$L$_{\rm Edd}$ values given in table 3 of Greene \& Ho (2004) were
corrected using the \mbh\ values from Barth et al. (2004), which are
better determined with higher quality data on H$\beta$. The L$_{\rm
bol}/$L$_{\rm Edd}$ values for the soft-X-ray selected samples are from
Grupe \etal (2004). One has to be cautious in comparing these samples
because they do not use the same prescription to estimate L$_{\rm
bol}$. Greene \& Ho (2004) use L$_{\rm bol} = 9.8\lambda$L$_{5100}$
uniformly for the entire sample, while Grupe \etal (2004) estimate the
bolometric correction separately for each object in the sample using the
observed spectral energy distribution. Inspection of figure 8 in Grupe
(2004) reveals that the two bolometric corrections are consistent with
each other. BH masses in all the three samples are estimated using
H$\beta$, so carry similar uncertainties. In figure 3 we plot the
distributions of L$_{\rm bol}/$L$_{\rm Edd}$ for the three samples. What
is seen is that the soft-X-ray selected NLS1 sample peaks at high
L$_{\rm bol}/$L$_{\rm Edd}$, the optically selected sample peaks at
lower L$_{\rm bol}/$L$_{\rm Edd}$, and the BLS1 sample peaks at even
lower L$_{\rm bol}/$L$_{\rm Edd}$. Indeed the mean $\log$L$_{\rm
bol}/$L$_{\rm Edd}$ of the soft-X-ray selected NLS1 sample is $+0.24$,
that of optically selected NLS1 sample is $-0.45$ and that of soft X-ray
selected BLS1s is $-0.75$ (the average L$_{\rm bol}/$L$_{\rm Edd}$ of
optically selected BLS1s may be even lower). This result is consistent
with Williams \etal (2004) who found that the mean value of soft X-ray
power-law slope $\Gamma$ decreases steadily from soft X-ray selected
NLS1s to optically selected NLS1s to BLS1s.

The above results imply that soft X-ray selected NLS1s are highly likely
to contain AGNs with large L$_{\rm bol}/$L$_{\rm Edd}$, but optically
selected NLS1s are not. Since AGNs with large L$_{\rm bol}/$L$_{\rm
Edd}$, and so large \dm, are the likely ones with rapidly growing BHs,
only they should have a distinct locus on the \mbh--\sig\ plane, away
from the Tremaine \etal relation. It should then be no surprise to find
that NLS1s with relatively smaller L$_{\rm bol}/$L$_{\rm Edd}$, notably
optically selected NLS1s, lie close to the Tremaine \etal relation, as
shown in, e.g., Barth \etal (2004).

\section{Discussion \& Conclusions}

Our result at face value, that highly accreting AGNs at low redshift
have lower \mbh/\sig$^4$ ratios than those of AGNs with low accretion rates,
has implications toward our understanding of the origin of the
\mbh--\sig\ relation (the AGNs with high L$_{\rm bol}/$L$_{\rm Edd}$ are
a subset of NLS1s while BLS1s have lower L$_{\rm bol}/$L$_{\rm
Edd}$). It tells us that BHs grow rapidly in their high accretion mode
and approach the \mbh--\sig\ relation asymptotically. This appears to be
the case at least at low redshift where BHs grow in well-formed bulges.

At higher redshifts, however, the situation appears to be different. As
discussed in \S 1, quasars at high redshift appear to lie above the
\mbh--\sig\ relation. This implies that the BHs in these quasars are
likely to have already grown to their ``final'' mass, but their host
galaxies have yet to grow further through interactions and mergers. Thus
the tracks of high redshift quasars on the \mbh--\sig\ plane may be
horizontal, from low to high mass bulges while that of low redshift
Seyfert galaxies may be vertical, from low to high BH masses. At low
redshift, where merger rates are very low, bulges in some galaxies have
grown to their ``final'' mass/ velocity dispersion, before their BHs
have fully grown. This is consistent with the newly emerging picture of
``downsizing of AGN activity'' or the ``anti-hierarchical BH growth'' in
which high mass BHs grow rapidly at high redshift while lower mass BHs
grow at successively lower redshifts (Merloni 2004). Since we are
probing BHs of $\sim 10^6$\msun\ in our NLS1 sample, it makes sense that
we catch them in their growing stage at present epoch (Mathur \& Grupe
2004).

Thus, we argue that there is no single answer to the origin of the
\mbh--\sig\ relation; it is a function of redshift. The feedback from
AGN may be the primary mechanism governing the co-evolution BHs and
galaxies at high redshift (Hopkins et al. 2005), but it may not work at
low redshift where merger rates are exceedingly low. Of course, not all
bulges are formed through mergers; disk/bar instability can also result
in formation of pseudo-bulges (Kormendy \& Kennicutt 2004). To our
knowledge there is no theoretical model on the co-evolution of BHs and
pseudo-bulges to explain the \mbh--\sig\ relation via feedback. Some
other mechanism, e.g. controlled accretion rate due to capture of bulge
stars by the accretion disk (Kollmeier \& Miralda-Escude 2004), may play
a dominant role at low redshift in ultimately placing all galaxies on
the \mbh--\sig\ relation.

On the other hand, our result at face value may not be correct due to
incorrect estimates of \sig\ (incorrect \mbh\ estimates are unlikely;
Paper I). It just may be that all galaxies, with dead or active BHs,
with low or high accretion rates, follow the same \mbh--\sig\
relation. Implications of such a result are perhaps even more
interesting. As shown in Paper I (their figure 2), the distributions of
BH masses of our soft X-ray selected NLS1 and BLS1 samples are
significantly different, with average $\log$\mbh\ of NLS1s being 6.9 and
that of BLS1s equal to 7.9. If all these galaxies lie on the \mbh--\sig\
relation, it would imply that NLS1s preferentially reside in host
galaxies with low mass/velocity dispersion bulges, or in later type
galaxies. Since a subset of these NLS1s have high \dm=\dM/\dMedd, this
implies that {\it BHs with high Eddington ratio preferentially reside in
later type galaxies} at low redshifts, compared to AGNs with low
Eddington accretion. This may be even a more challenging theoretical
problem to address, than the origin of the \mbh--\sig\ relation, for
which there seem to have plenty of explanations. Some of this might
simply be a gas supply issue. If the gas available to feed the BHs at
present epoch is relatively small, the accretion rate on smaller mass
BHs may be close to Eddington, but it would be substantially
sub-Eddington for higher mass BHs. When we search for luminous NLS1s
based on their narrow H$\beta$ widths, we are necessarily looking for
smaller mass BHs with high accretion rates, and they are to be found in
later type galaxies. Both the high mass and low mass BHs would be
growing in mass by the same rate then, but the fractional growth would
still be higher for the lower mass BHs. To keep the high \dm\ BHs
from moving away from the \mbh--\sig\ relation, the gas supply in late
type galaxies will have to be lower. Finding the locus of highly
accreting AGNs on the \mbh--\sig\ plane is important either way.

We are grateful to the anonymous referee for careful and thoughtful
reading of the paper and useful comments. We also thank Greene \& Ho for
their comments.


\begin{figure}
\psfig{file=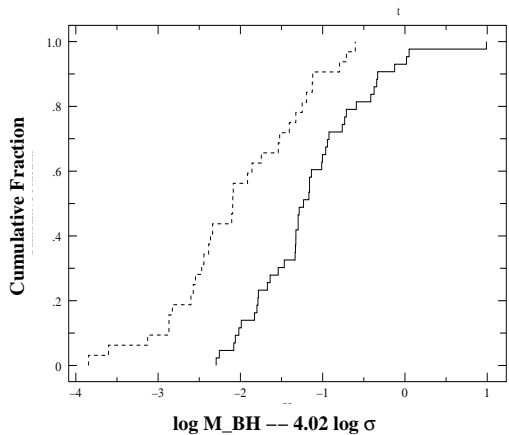,width=3in,height=2.5in}
\caption{Cumulative fraction of a K-S test for the distribution of
 $\log$\mbh$-4.02\log$\sigo\ for the two populations of NLS1s (dashed)
 and BLS1s (solid). The data from Grupe \& Mathur (2004) are modified with new
 values of \sigo\ calculated assuming equation 1. That the two
 populations are different remains a robust statistical result.}
\end{figure}

\begin{figure}
\psfig{file=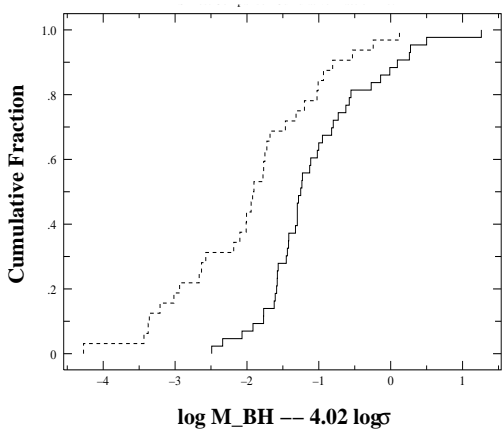,width=3in,height=2.5in}
\caption{Same as Figure 1, but with \sigo\ in which the blue wing of
 the [OIII] line is removed. The two populations are still significantly
 different.}
\end{figure}


\begin{figure}
\psfig{file=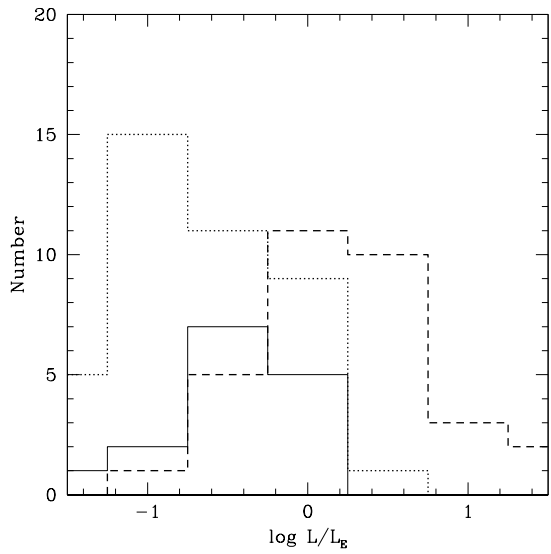,width=3in}
\caption{The distributions of
L$_{\rm bol}/$L$_{\rm Edd}$ for three samples: soft X-ray selected NLS1s
from Grupe \& Mathur (dashed), optically selected NLS1s from Barth et
al. (solid), and soft X-ray selected BLS1s from Grupe \& Mathur
(dotted). Given the large uncertainties in estimating L$_{\rm
bol}/$L$_{\rm Edd}$, the trend in the distributions of the three samples
is of interest, not their exact values. }
\end{figure}

\clearpage



\newpage
\begin{thebibliography}{}
\bibitem[Barth et al. 2005]{barth05} Barth, A., Greene, J., \& Ho,
 L. 2005, ApJL, 619, 151
\bibitem[Boroson 2003]{bor03} Boroson, T.A., 2003, \apj, 585, 647
\bibitem[]{} Corbett, E.A. et al. 2003, MNRAS, 343,705
\bibitem[Ferrarese \& Merritt 2000]{ferr00} Ferrarese, L., \& Merritt, D.,
2000, \apj, 539, L9
\bibitem[Ferrarese et al. 2001]{ferr01} Ferrarese, L., Pogge, R.W., Peterson,
B.M., Merritt, D., Wandel, A., \& Joseph, C.L., 2001, \apj, 555, L55
\bibitem[Gebhardt et al. 2000a]{geb00a} Gebhardt, K., Bender, R., Bower, G.,
Dressler, A., Faber, S.M., et al., 2000, \aap, 539, L13
\bibitem[Gebhardt et al. 2000b]{geb00b} Gebhardt, K., Kormendy, J., Ho, L.C.,
Bender, R., Bower, G., et al., 2000, \apj, 543, L5
\bibitem[Greene \& Ho 2004]{gh04} Greene, J. \& Ho., L. 2004 ApJ, 610, 722
\bibitem[Greene \& Ho 2005]{gh05} Greene, J. \& Ho., L. 2005 ApJ, in press. astro-ph/0503675
\bibitem[Grupe et al. 2004]{gru03a} Grupe, D., Wills, B.J., Leighly, K.M., \&
Meusinger, H., 2004, \aj, 127, 156
\bibitem[Grupe 2004]{gru03b} Grupe, D., 2004, \aj, 127, 1799
\bibitem[Grupe \& Mathur 2004]{gm04} Grupe, D., \& Mathur, S. 2004,
 ApJL, 606, 41
\bibitem[Hopkins et al. 2005]{} Hopkins, P. et al. 2005, ApJ, submitted. 
                                  astro-ph/0504253
\bibitem[Kaspi et al. 2000]{kas00} Kaspi, S., Smith, P.S., Netzer, H., Moaz,
D., Jannuzi, B.T., \& Giveon, U., 2000, \apj, 533, 631
\bibitem[King \& Pounds 2003]{king03} King, A.R., \& Pounds, K.A., 2003,
\mnras, in press, astro-ph/0305541
\bibitem[Kollmeier \& Miralda-Escude 2005]{} Kollmeier, J. \& Miralda-Escude, J., 2005, ApJ, 619, 30
\bibitem[]{} Kormendy, J. \& Kennicutt, R. 2004, ARAA, 42, 603
\bibitem[]{} Lu, Y. \& Yu, Q. 1999 astro-ph/9911289
\bibitem[Mathur2000]{mat00} Mathur, S., 2000, \mnras, 314, L17
\bibitem[Mathur et al. 2001]{mat01} Mathur, S., Kuraszkiewicz, J., \& Czerny,
B., 2001, New Astronomy, Vol. 6, p321
\bibitem[Mathur \& Grupe 2004]{} Mathur, S. \& Grupe, D. 2004, in
 ``Growing Black Holes", Ed. A. Merloni, S. Nayakshin and R. Sunyaev,
 Springer-Verlag series of "ESO Astrophysics Symposia"
\bibitem[Mathur \& Grupe 2005]{} Mathur, S. \& Grupe, D. 2005, A\&A, in press.
\bibitem[McLure \& Dunlop 2002]{mclure02} 
        McLure, R.J., \& Dunlop, J.S., 2002, \mnras, 331, 795
\bibitem[Merloni 2004]{} Merloni, A. 2004 in ``Growing Black Holes",
 Ed. A. Merloni, S. Nayakshin and R. Sunyaev, Springer-Verlag series of
 "ESO Astrophysics Symposia"
\bibitem[Merritt \& Ferrarese 2001]{merr01} Merritt, D., \& Ferrarese, L.,
2001, \apj, 547, 140
\bibitem[Nelson 2000]{nel00} Nelson, C.H., 2000, \apj, 544, L91
\bibitem[Nelson \& Whittle 1995]{nel95} Nelson, C.H., \& Whittle, M., 1995,
\apjs, 99, 67
\bibitem[Nelson \& Whittle 1996]{nel96} Nelson, C.H., \& Whittle, M., 1996,
\apj, 465, 96
\bibitem[Peterson et al. 2000]{pet00} Peterson, B.M., McHardy, I.M., Wilkes,
B.J., Berlind, P., Bertram, R., et al., 2000, \apj, 542, 161
\bibitem[Peterson 1993]{pet93} Peterson, B.M., 1993, PASP, 105, 247
\bibitem[Pounds et al. 1995]{pound95} Pounds, K.A., Done, C., \& Osborne, J.,
1995, \mnras, 277, L5
\bibitem[Rix et al. 1999]{rix99} Rix, H-W. et al. 1999 astro-ph/9910190
\bibitem[Shields et al. 2003]{shi03} Shields, G.A., Gebhardt, K., Salviander,
S., Wills, B.J., Xie, B., Brotherton, M.S., Yuan, J., \& Dietrich, M., 2003,
\apj, 583, 124
\bibitem[Tremaine et al. 2003]{tre02} Tremaine, S., Gebhardt, K., Bender, R.,
et al., 2003, \apj, 574, 740
\bibitem[]{} Whittle, M. 1985, MNRAS, 213, 33
\bibitem[]{} Whittle, M. 1992, ApJ, 387, 121
\bibitem[Williams et al. 2002]{williams02} Williams, R.J., Pogge, R.W., \&
Mathur, S., 2002, \aj, 124, 3042
\bibitem[Williams et al. 2004]{williams04} Williams, R.J., Mathur,
 S., \& Pogge, R.W. 2004, ApJ, 610, 737
\end{thebibliography}
\end{document}